# Thermodynamics, Dynamics, and Kinetics of Nanostructured Fluid-Solid Interfaces

Carlos E. Colosqui
Department of Mechanical Engineering, Stony Brook University.
Department of Applied Mathematics & Statistics, Stony Brook University.
E-mail: carlos.colosqui@stonybrook.edu

**Abstract**

This article covers thermodynamic, dynamic, and kinetic models that are suitable for the analysis of wetting, adsorption, and related interfacial phenomena in colloidal and multiphase systems. Particular emphasis is made on describing crucial physical assumptions and the validity range of the described theoretical approaches and predictive models. The classical sharp interface treatment of thermodynamic systems where a perfectly smooth surface is assumed to separate homogeneous phases can present significant limitations when analyzing systems that are subject to thermal motion and present multiple metastable states caused by interfacial heterogeneities of nanoscale dimensions. Mesoscopic approaches such as stochastic Langevin dynamics can extend the application of sharp interface models to a wide variety of systems exhibiting metastability as they undergo thermal motion. For such metastable systems, dynamic and kinetic equations can describe the evolution of observable (macroscopic) variables as the system approaches thermodynamic equilibrium. Sufficiently close to equilibrium, Kramers theory of thermally activated escape from metastable states can be effectively employed to describe diverse wetting and interfacial processes via kinetic equations. Future directions for further advancement and application of thermodynamic, dynamic, and kinetic models are briefly discussed in the context of current technological developments involving nanoparticles, nanofluidics, and nanostructured surfaces.

**Nomenclature**

$A_d$ $[\mathrm{m}^2]$: Characteristic projected area of a surface heterogeneity or defect.

$A_{ij}$ $[\mathrm{m}^2]$: Surface area of the interface between the $i$-th and $j$-th phase

$D$ $\left[\frac{\mathrm{Nm}}{\mathrm{s}}\right]$: Rayleigh dissipation function

$E = K + P$ $\left[\frac{\mathrm{Nm}}{\mathrm{s}}\right]$: Total system energy

$F$ [Nm]: Helmholtz free energy

$f_i^c$ [N]: Resulting conservative force acting on the $i$-th system variable $q_i$

$f_i^d$ [N]: Resulting dissipartive force acting on the $i$-th system variable $q_i$

$K$ [Nm]: Total kinetic energy of the system

$k_B = 1.3806485 \times 10^{-23}$ $\left[\frac{\mathrm{Nm}}{\mathrm{K}}\right]$: The Boltzmann constant

$\ell$ [m]: Mean contact line position

$\ell_C$ [m]: Capillary length

$m$ [kg]: Molecular mass

$N_i^{(k)}$ [#mol]: Number of molecules of the $k$-th component in the $i$-th phase

$P$ [Nm]: Total potential energy of the system

$p_i$ $\left[\frac{\mathrm{N}}{\mathrm{m}^2}\right]$: Pressure in the $i$-th phase

$q_\mathrm{i}$ : The $i$-th system variable or generalized coordinate of a finite set $\{q_i\}$ $(i = 1, N_q)$

$R$ [m]: The radius of a spherical particle or the contact area of a hemispherical droplet

$R_C$ [m]: The local radius of curvature

$S$ $\left[\frac{\mathrm{Nm}}{\mathrm{K}}\right]$: Entropy

$T$ [K]: Absolute temperature

$U$ [Nm]: Internal energy

$V_i$ [m$^3$]: Volume occupied by the $i$-th phase

$\Gamma_\pm$ $\left[\frac{1}{\mathrm{s}}\right]$: Rate of escape from a metastable state in the forward/backward (+/-) direction

$\gamma_{ij}$ $\left[\frac{\mathrm{N}}{\mathrm{m}}\right]$: Surface energy (or interfacial tension) of the interface between the $i$-th and $j$-th phase

$\eta$ $\left[\frac{\mathrm{Ns}}{\mathrm{m}^2}\right]$: Molecular or shear viscosity

$\theta$ [rad]: The macroscopic (observed) contact angle

$\theta_Y$ [rad]: The Young contact angle

$\lambda_M$ [m]: Characteristic size of molecular adsorption sites on a solid surface

$\mu_i^{(k)}$ $\left[\frac{\mathrm{Nm}}{\#\mathrm{mol}}\right]$: Chemical potential of the $k$-th component in the $i$-th phase (energy per molecule)

$\nu$ $\left[\frac{1}{\mathrm{m}^3}\right]$: Molecular volume

$\xi_i$ $\left[\frac{\mathrm{Ns}}{\mathrm{m}}\right]$: Damping coefficient for the variable $q_i$

$\sigma$ [m]: Molecular diameter

$\tau$ [N]: Line tension at a three-phase contact line

$\tau_m$ [s]: Microscopic relaxation time

$\tau_M$ [s]: Macroscopic relaxation time

$\Omega$ [Nm]: Grand thermodynamic potential (Landau free energy)

# Introduction

A wide variety of natural and industrial processes that are essential to modern technologies involve the wetting of solid surfaces by simple and complex fluids as well as the adsorption and adhesion of colloidal particles (e.g., micro/nanoscale beads, droplets, bubbles, macromolecules) to liquid-fluid and fluid-solid interfaces or membranes.[1-3] Our current fundamental understanding of the behavior of liquids and colloids at interfaces has helped to develop applications ranging from self-assembly of nanomaterials [4, 5] and additive manufacturing[6, 7] to drug delivery[8, 9] and water treatment,[10, 11] among many others. Predicting the dynamics of wetting and adsorption at interfaces requires not only understanding mechanical and hydrodynamic effects, which can be effectively described by conventional continuum-based models, but also nanoscale phenomena, such as intermolecular and surface forces and Brownian motion, that require careful modeling when adopting continuum descriptions. Considering the complexity of the numerous intermolecular processes that give rise to (isotropic and homogeneous) macroscopic bulk behavior of fluids and colloids, one can expect significant challenges in modeling the macroscopic interfacial behavior when physico-chemical anisotropies and heterogeneities are caused by the presence of liquid-fluid and fluid-solid interfaces.

Models based on continuum thermodynamics of interfaces [12-14] (e.g., Young-Dupre and Young-Laplace equation) have been extensively adopted with varying degrees of success. Such conventional continuum-based descriptions consider the interface between two phases as a sharp and sufficiently smooth surface that can be described using differential geometry. These classical descriptions have effectively rationalized the equilibrium behavior and some fundamental dynamic aspects of wetting and adhesion in applications ranging from self-assembly of microparticles at interfaces [15-18] and Pickering emulsions [19, 20] to spontaneous spreading [21-23] and capillary imbibition.[24-26] In recent decades, with the advent of nanofabrication and advanced characterization techniques, researchers have found some significant limitations of classical continuum descriptions for predicting the dynamic behavior of diverse micro- and nanoscale systems. This article describes the core of classical theories and some recent efforts to advance our understanding of the dynamics and kinetics of wetting, adsorption, and adhesion at liquid-fluid and fluid-solid interfaces. In particular, this article describes models that attempt to better account for the effects of nanoscale physico-chemical surface features of random or synthetic nature, the effect of finite-range molecular interactions, and thermal motion.

# Thermodynamics of Sharp Interfaces

In this section we will consider the interfacial region between two phases as a sharp and smooth "dividing surface" following Gibbs' original treatment of the thermodynamics of interfaces. [27] The fundamental relations presented in this section can be obtained from more detailed thermodynamic descriptions considering that an interface is a thin but finite region, where local properties change gradually, and thus has its own entropy and chemical potential. Despite significant physical simplifications, sharp interface descriptions have been effectively employed for the ultimate purpose of predicting equilibrium states for diverse wetting problems including droplet spreading and particle adsorption at interfaces.

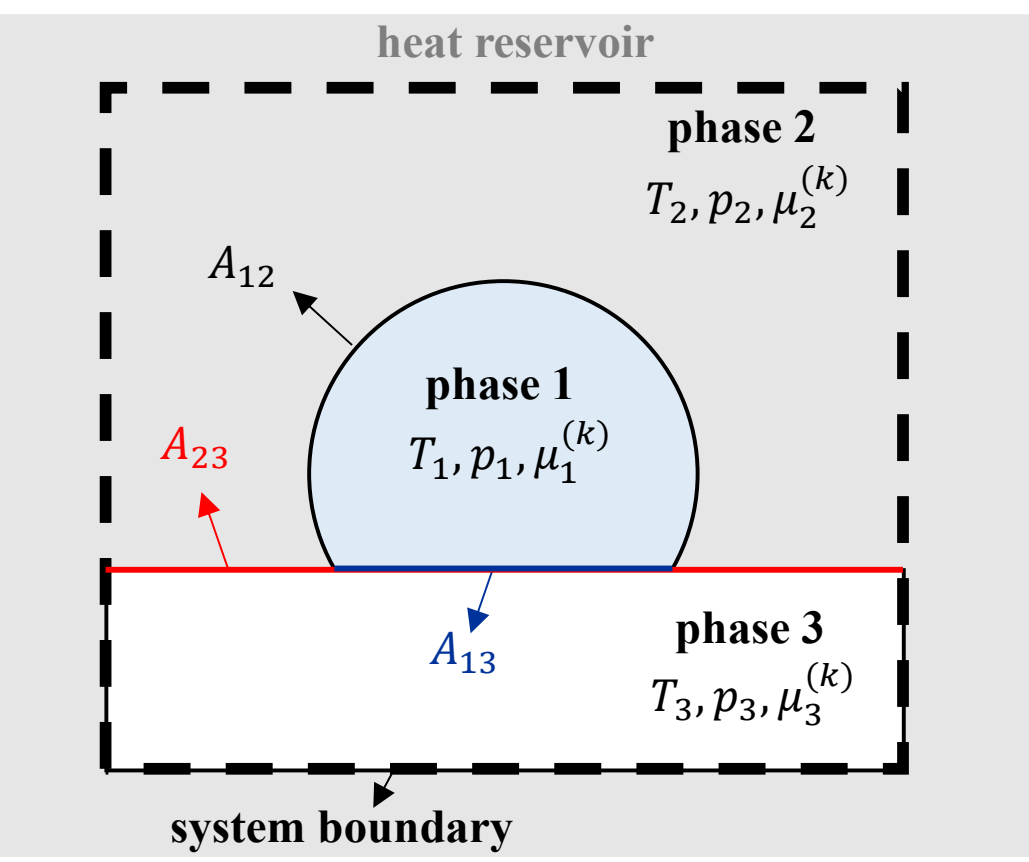

Figure 1: Sharp interface description of an open thermodynamic system of volume $V$. The system is composed of an arbitrary number of chemical species ($k = 1, N_s$) and three phases ($i = 1,3$) separated by sharp interfaces with surface areas $A_{12}$, $A_{13}$, and $A_{23}$. Thermodynamic potentials to study the illustrated system are given in eqns [1]-[3].

We begin our analysis by considering an open thermodynamic system of volume $V$ that consists of three homogeneous phases ($i = 1,3$) in thermodynamic equilibrium that are separated by sharp interfaces with surface areas $A_{ij}$ [See Figure 1]. The studied system is composed of a number $N_i^{(k)}$ of molecules[1] of substance $k$ ($k = 1, N_s$) that can occupy the $i$-th phase and can exchange mass with a much larger heat reservoir at temperature $T$. Under thermodynamic equilibrium all phases in the system must have the same temperature $T_i = T$ and chemical potentials $\mu_i^{(k)} = \mu^{(k)}$ but not neccesarily the same mechanical pressures (i.e., $p_i - p_j \gtreqless 0$) since mechanical equilibrium involves forces due to deformation of the interfaces. Under the proposed idealization, sharp interfaces occupy no physical volume (i.e., $dV = \sum dV_i$) and increasing their surface areas requires a specific energy per unit area $\gamma_{ij}$; this quantity is known as the surface tension. A differential change in internal energy is given by

$$dU = TdS - \sum p_i dV_i + \sum \gamma_{ij} dA_{ij} + \sum \mu^{(k)} dN_i^{(k)}. \qquad (1)$$

Hereafter, the summations are carried over each phase ($i = 1,3$) and substance ($n = 1, N_s$), and each of the three different interfaces $A_{12}, A_{13}, A_{23}$. For thermodynamic systems evolving at constant volume and temperature ($dV = 0, dT = 0$) it is convenient to employ the Helmholtz free energy $F = U - TS$, for which differential changes are given by $dF = dU - TdS - SdT$, and thus we have

$$dF = -\sum p_i dV_i + \sum \gamma_{ij} dA_{ij} + \sum \mu^{(k)} dN_i^{(k)}. \qquad (2)$$

Furthermore, when the studied system evolves at constant temperature, volume, and chemical potential ($dV = 0, dT = 0, d\mu^{(k)} = 0$) it is convenient to employ the grand potential $\Omega = U - TS - \sum \mu^{(k)} dN_i^{(k)}$ for which the differential changes are given by

$$d\Omega = -\sum p_i dV_i + \sum \gamma_{ij} dA_{ij} . (3)$$

It is worth noticing that while changes in the Helmholtz free energy (eqn 2) give the reversible work performed by a closed system, changes in the grand potential (eqn 3) give the reversible work performed by an open system. Hence, the Helmholtz free energy F and grand potential Ω, also known as the Landau free energy, can determine conservative forces when modeling the system dynamics.

---

[1] In this article the amount of $k$-th substance $N^{(k)}$ is given in number of molecules. Whether the amount of substance is measured in number of molecules or moles prescribes that chemical potentials $\mu^{(k)}$ must be given in units of J/molecule or J/mol.

### Line Tension and excess energy at the three-phase contact line.

The expressions in eqns (1)-(3) neglect energy contributions $\mathrm{d}U_L = \tau dL$ due to changes in the perimeter $L$ of the three-phase contact line, which are proportional to the so-called line tension $\tau$. [28, 29] The line tension is the 1D analog of the surface tension and it accounts for the "excess energy" at the contact line where molecules interact with all three different phases. A basic estimation of the line tension magnitude for simple liquids with molecular diameters $\sigma \sim 0.1$ nm gives $|\tau| \simeq k_B T/\sigma \sim 10^{-11}$ N. Detailed theoretical estimations for simple molecular liquids give $|\tau| \sim 10^{-12}$ to $10^{-10}$ N and experimental studies for different systems (e.g., droplets, emulsions, foams) report positive and negative values with magnitudes $|\tau| \sim 10^{-8}$ to $10^{-6}$ N. [30] For the case of simple fluids where typical surface tensions values are $\gamma \sim 10^{-2}$ N/m$^2$ one can estimate that line tension contributions to the system energy must be considered for characteristic system dimensions $\ell < \tau/\gamma \sim 1$ to 10 nm. Unfortunately, determining the line tension for a given liquid pair and a solid surface can be a challenging task and a matter open to debate. [31, 32]

## Thermodynamic Equilibrium

Adopting a sharp interface model where physical properties are uniform within each phase of the system, eqns (1)-(3) can be readily applied to determine the conditions for thermodynamic equilibrium, where the first-order energy variation must vanish $\delta U = 0$. For a set of $N_q$ independent macroscopic (observable) variables $\{q_n\}$ $(n = 1, N_q)$ that parametrize the system energy $U = U(\{q_n\})$, finding equilibrium conditions requires solving the set of $N_q$ independent equations $\partial U/\partial q_i = 0$. The set of macroscopic variables $\{q_n\}$ can include state variables (e.g., pressure, volume, interfacial areas) and/or geometric parameters, as we shall see for specific problems in the following sections. Given specific geometric configurations [Figure 2], applying eqns (1)-(3) enables a rigorous derivation of well-known relations such as the Young-Laplace equation, relating interfacial curvature and pressures in each phase, and the Young equation for determining the equilibrium contact angle.

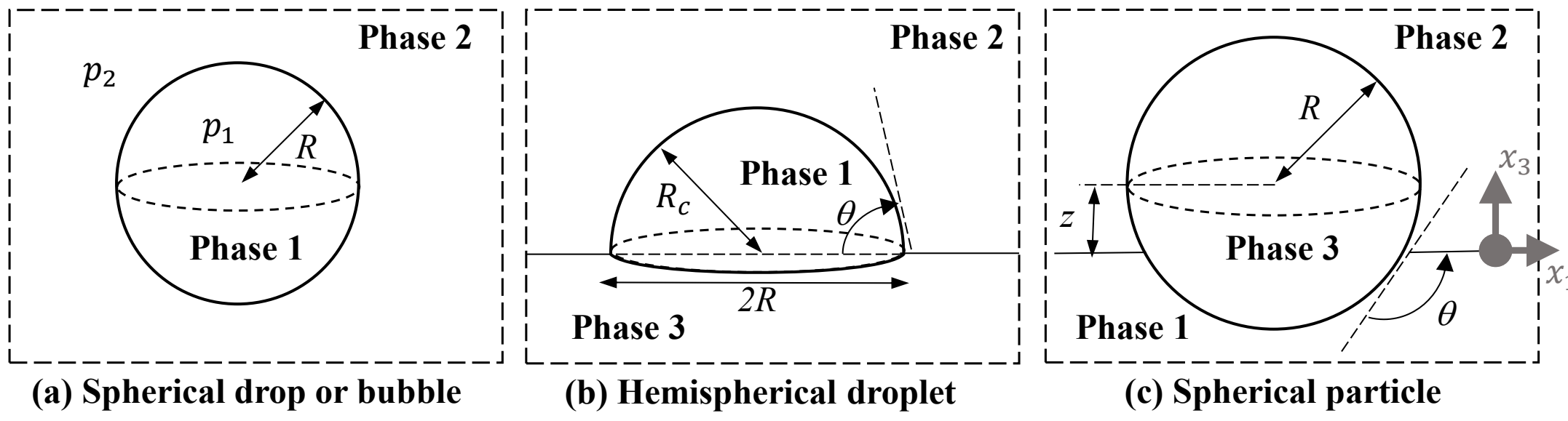

Figure 2: Different interfacial configurations where a sharp interface treatment can effectively predict equilibrium values of geometric parameters. (a) Spherical drop or bubbles. (b) Hemispherical droplet or bubble spreading on flat surface. (c) Spherical particle at a flat liquid-fluid interface.

### Droplets & Bubbles: Laplace & Kelvin Equations

Assuming that a droplet or bubble in thermodynamic equilibrium with the surrounding ambient phase [See Figure 2a] has a perfectly spherical shape, one can parametrize its volume $V_1 = (4/3)\pi R^3$ and surface area $A_{12} = dV_1/dR = 4\pi R^2$ using the radius $R$ as the only variable. For a fixed system volume $(dV_1 = -dV_2)$, constant temperature, and chemical potential, it is convenient to work with the grand potential $\Omega$ to find equilibrium conditions. For a spherical droplet or bubble, eqn (3) gives $d\Omega/dR = -(p_1 - p_2)dV_1/dR + \gamma_{12}\, dA_{12}/dR = 0$, which leads to the well know Laplace pressure equation

$$p_1 - p_2 = \gamma_{12} \frac{dA_{12}/dR}{dV_1/dR} = \gamma_{12} \frac{2}{R} \quad (4)$$

relating the pressure difference between phases with the interfacial tension and droplet radius. It is worth noticing that eqn (4) is not sufficient to determine the actual value of the equilibrium pressures in each phase. For that purpose, one needs to invoke the equality of chemical potentials $\mu_1^{(k)} = \mu_2^{(k)} = \mu^{(k)}$. For a single-component (incompressible) liquid droplet ($V_1/N_1 = \text{const.}$) surrounded by a vapor phase obeying the ideal gas law $p_2V_2 = N_2k_BT$ we have $\mu_i(p_i,T) = \mu_i(p_0,T) + \int_{p_0}^{p_i}(V_i/N_i)dp$, where $p_0$ is the pressure at a reference state, and invoking eqn (4) one can readily obtain the relation

$$(p_1 - p_0)\left(\frac{V_1}{N_1}\right) = k_BT\, ln\left(\frac{p_1}{p_0} + \frac{2\gamma_{12}}{Rp_0}\right). (5)$$

Conversely, for a compressible vapor bubble surrounded by a liquid phase we obtain

$$k_BT\, ln\left(\frac{p_1}{p_0}\right) = (p_1 + 2\gamma_{12} - p_0)\left(\frac{V_2}{N_2}\right), (6)$$

where $p_0$ is commonly chosen to be the saturated vapor pressure at the system temperature $T$. The relation in eqns (5)-(6) are commonly known as the Kelvin equation and can be extended to multicomponent systems. [12, 33] These types of equations provide the basis to analyze the equilibrium and dynamic behavior of various interfacial phenomena such as nucleation, capillary condensation, or Ostwald ripening. [34-36]

## The Equilibrium Contact Angle: Young's Law

A fundamental element in analytical descriptions of wetting and interfacial phenomena is the concept of the macroscopic contact angle $\theta$, which is the observable angle between a liquid-fluid interface and either one of the solid-fluid interfaces [see Figure 2b-c]. According to Young's law [37] the value of the contact angle in thermodynamic equilibrium is $\theta = \theta_Y$, where the so-called Young contact angle $\theta_Y$ is given by the relation

$$cos\, \theta_Y = \frac{\gamma_{23} - \gamma_{13}}{\gamma_{12}}. (7)$$

Here, the contact angles $\theta$ and $\theta_Y$ are measured in phase 1 and phase 3 correspond to the solid surface. Neglecting the effect of external fields (e.g., gravitational or electrostatic fields) acting in arbitrary directions, eqn (7) establishes that the equilibrium contact angle is independent of the system dimensions. Under certain general assumptions it can be further established that Young's law is valid when external fields act normal to the solid surface.[38] However, size dependence of the equilibrium contact angle has been reported in systems with small dimensions. [39-41] For such systems Young's law has been extended to include line tension effects, in which case eqn (7) becomes $\cos\theta_Y = (\gamma_{23} - \gamma_{13})/\gamma_{12} - \tau/\gamma_{12}R_c$, where $R_c$ is the radius of curvature of the contact line.

Although Young's law can be readily obtained on a perfectly flat and chemically homogeneous surface via simple mechanistic arguments, it is important to understand the specific physical and geometric assumptions under which $\theta_Y$ gives an accurate estimate of the actual equilibrium value of the thermodynamic variable $\theta$. In particular, physico-chemical surface heterogeneities with dimensions ranging from nanometers to microns can induce nontrivial effects such as hysteresis of the equilibrium contact angle [42-44] and metastable wetting states (e.g., Wenzel and Cassie states).[45-47] In the following sections we will analyze the specific cases of a hemispherical droplet on a perfectly smooth surface and a spherical particle on a flat liquid-fluid interface and determine equilibrium conditions using the thermodynamic relations in eqns (1)-(3). Later sections will consider the presence of nanoscale heterogeneities on the liquid-solid interface and their effects on the equilibrium and dynamic behaviors.

## Hemispherical Droplet on a Solid Surface

Similar thermodynamic analysis employed for drops and bubbles fully surrounded by a fluid can be applied to the case of a liquid droplet or bubble sitting on a flat solid surface [See Figure 2b]. Neglecting gravitational effects and the action of other external forces (e.g., van de Waals, electrostatic forces) the droplet is expected to become a hemispherical cap in order to minimize its surface energy. In practice, the hemispherical shape assumption is valid for droplets with dimensions smaller than the capillary length $\ell_C$ (e.g., about 3 mm for water-air systems) and larger than the interaction range of surface forces (e.g., usually 10 to 100 nm for colloidal systems).

The volume of the hemispherical droplet $V_1(R,\theta) = R^3 f_V(\theta)$ is prescribed by the contact radius $R$ and contact angle $\theta$; here, $f_V(\theta) = \pi[2/3 - (3/4)\cos\theta + (1/12)\cos 3\theta]/\sin^3\theta$. Furthermore, the interfacial areas for a hemispherical cap are $A_{13}(R,\theta) = \pi R^2$ and $A_{12}(R,\theta) = 2\pi R^2/(1+\cos\theta)$, and thus we can cast eqn (3) as $d\Omega(R,\theta) = (\partial\Omega/\partial R)dR + (\partial\Omega/\partial\theta)d\theta$. In general, the equilibrium contact radius and contact angle can be found by solving for $\partial\Omega/\partial R = 0$ and $\partial\Omega/\partial\theta = 0$ in order to satisfy the thermodynamic equilibrium condition $\delta\Omega = 0$. Assuming hemispherical cap of known volume $V_1$ the contact radius and contact angle are not independent and we can define $\theta(R) = f_V^{-1}(V_1/R^3)$. For the particular case of a non-evaporating and incompressible droplet we have $dV_1 = 0$ and introducing the Young contact angle (eqn 7), the grand potential

$$\Omega(R) = \gamma_{12}\pi R^2\left(\frac{2}{1+\cos\theta(R)} - \cos\theta_Y\right) + C \quad (8)$$

can be parametrized by the contact radius $R$; here, $C = \Omega(0)$ is an arbitrary additive constant. The constant volume constraint imposes the relation $d\theta/dR = -(\partial V_1/\partial R)/(\partial V_1/\partial\theta) = 3f_V/R\dot{f_V}$ and thus changes in the grand potential are given by

$$d\Omega = -2\pi R\gamma_{12}(\cos\theta(R) - \cos\theta_Y)dR. \quad (9)$$

From eqn (9) we readily find that thermodynamic equilibrium is attained for $\theta = \theta_Y$ and $R = [V_1/f_V(\theta_Y)]^{1/3}$. We have thus verified that the equilibrium contact angle is given by Young's law (eqn 7) for the case of an incompressible liquid droplet of hemispherical shape that spreads on a perfectly flat surface when the effects of external fields are negligible.

## Spherical Particle at a Flat Interface

We continue to employ the thermodynamic analysis described for droplets and bubbles for the case of a rigid spherical particle of radius $R$ that straddles a perfectly flat interface located at position $x_3 = 0$ as illustrated in Figure 2c. The assumption that the surface remains perfectly flat when breached by the particle can only be justified for particle sizes smaller than the capillary length ($R \ll \ell_C$) and larger than the range of interaction of surface forces ($R \gtrsim 10$ to 100 nm). When the contact angle $\theta$ is measured on the phase-1 side [see Figure 2c] the center-of-mass of the particle is located at a distance

$$z = -R\cos\theta \quad (10)$$

from the flat interface. Under the adopted geometric assumptions, the interfacial areas $A_{12}(z) = \pi R^2\sqrt{1-(z/R)^2}$, $A_{13}(z) = 2\pi R^2(1 - z/R)$, and $A_{23}(z) = 4\pi R^2 - A_{13}$ can be determined as a function of the particle position $z$. In accordance with Laplace's law (eqn 4) the pressures on each side of the flat interface must be equal and constant ($p_1 = p_2 =$ const.); this assumption is justified in detail via minimization of energy in the following section. Changes in the grand potential (eqn 3) are thus given by $d\Omega = \gamma_{12}(dA_{12} - \cos\theta_Y\, dA_{13})$ and integration with respect to the particle position $z$ yields

$$\Omega(z) = \pi\gamma_{12}(z - z_E)^2 + C, \quad (11)$$

where $z_E = -R\cos\theta_Y$ is the particle position at thermodynamic equilibrium ($d\Omega/dz = 0$) and $C$ is an arbitrary constant. Hence, we find that for a spherical particle at a perfectly flat interface the equilibrium contact angle $\theta = \theta_Y$ is given by Young's law (eqn 7).

## Arbitrarily Curved Interfaces: The Young-Laplace Equation

When the shape of the studied phases is known, or assumed to be known (e.g., perfectly spherical droplets), thermodynamic potentials in eqns (1)-(3) can be readily expressed as a function of geometric parameters (e.g., contact radius/angle, particle position). For the more general case of an arbitrarily curved interface between two fluid phases [see Figure 3], the unknown shape $h = f(x_1, x_2)$ of the free surface under thermodynamic equilibrium conditions must be determined by minimizing the total system energy.

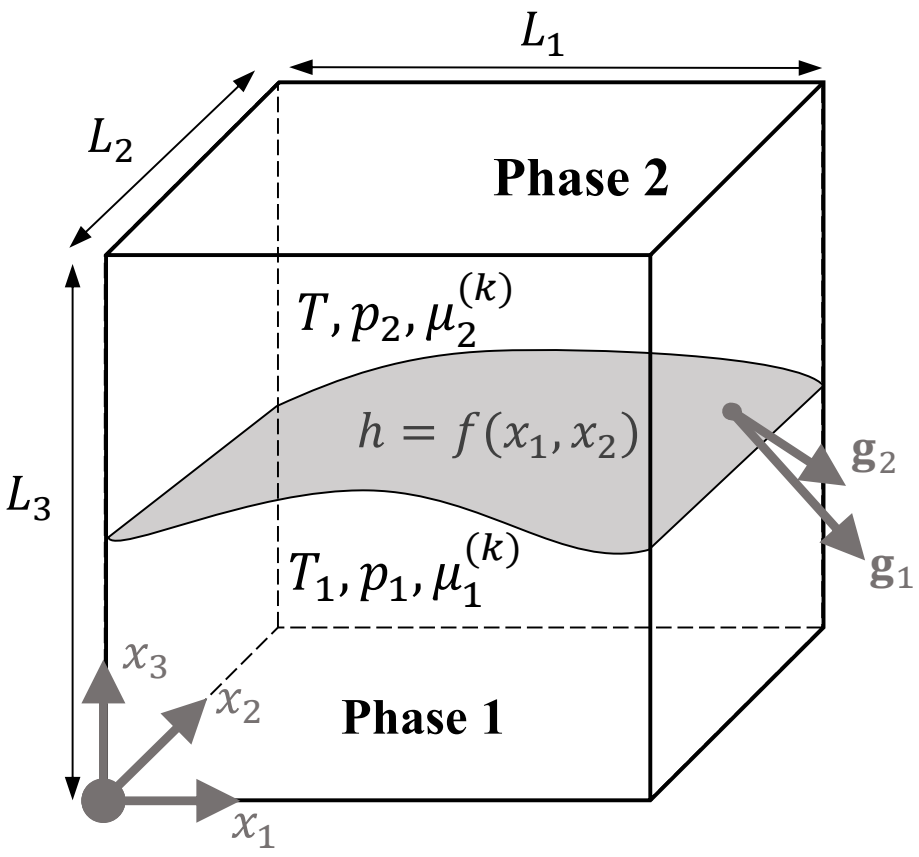

Figure 3: Two-phase system separated by a sharp interface at $h = f(x_1, x_2)$ that is arbitrarily curved. The system is assumed to be in thermal and chemical equilibrium. Local pressures $p_i(x_1, x_2, h)$ measured in each phase satisfy hydrostatic equilibrium conditions when external forces $g_i$ re applied on each side the interface.

We will analyze a system consisting of a rectangular cuboid with fixed length $L_1$, width $L_2$, and height $L_3$ [see Figure 3] where the two fluid phases are in thermal and chemical equilibrium ($T_i = T$, $\mu_i^{(k)} = \mu^{(k)}$). We thus aim to minimize the grand potential

$$\Omega[f] = \int_0^{L_1}\int_0^{L_2} \mathcal{P}\left(x_i, f, \dot{f}_i\right) dx_1 dx_2\,, (12)$$

defined as functional of $f$, where $\dot{f}_i \equiv \partial f/\partial x_i$ ($i = 1{,}2$) and

$$\mathcal{P}(x_i, f, \dot{f}_i) = -\int_0^f p_1\, dx_3 - \int_f^{L_3} p_2\, dx_3 + \gamma_{12}\sqrt{1 + \dot{f}_1^2 + \dot{f}_2^2}. (13)$$

To find the interface shape $f$ that minimizes $\Omega[f]$ one can solve the Euler-Lagrange equation[48]

$$\frac{\partial \mathcal{P}}{\partial f} = \frac{\partial}{\partial x_1}\left(\frac{\partial \mathcal{P}}{\partial \dot{f}_1}\right) + \frac{\partial}{\partial x_2}\left(\frac{\partial \mathcal{P}}{\partial \dot{f}_2}\right), (14)$$

which gives the expression commonly known as the Young-Laplace equation[49, 50]

$$\Delta p(x_1, x_2, h) = -\gamma_{12}\left(\frac{\partial}{\partial x_1}\frac{\dot{f}_1}{\sqrt{1 + \dot{f}_1^2 + \dot{f}_2^2}} + \frac{\partial}{\partial x_2}\frac{\dot{f}_2}{\sqrt{1 + \dot{f}_1^2 + \dot{f}_2^2}}\right), (15)$$

relating the pressure difference $\Delta p = p_1 - p_2$ between phases and the shape $f(x_1, x_2)$ of an arbitrarily curved interface. It is worth noticing that as result of adopting a sharp interface treatment, $p_1(x_1, x_2, h)$ and $p_2(x_1, x_2, h)$ are "bulk" pressures measured at the local interface position $(x_1, x_2, h)$ on the phase-1 and phase-2 side, respectively. Mechanical equilibrium in the $i$-th fluid phase requires that bulk pressures $p_i$ satisfy the hydrostatic equation $\nabla p_i = \rho_i \mathbf{g}_i$, where $\rho_i$ is the mass (or charge) density and $\mathbf{g}_i = \boldsymbol{\nabla}\phi_i$ is the net body (or electrostatic) force due to external fields $\phi_i$ in the $i$-th phase. Hence, eqn (15) can readily incorporate the effect of gravitational, electrostatic, or surface forces (e.g., van der Waals forces) when equilibrium bulk pressures defining $\Delta p$ account for external fields $\phi_i(x_1, x_2, h)$ on each side of the interface.

### Surface Heterogeneities and Roughness: Ideal vs Real Surfaces

As discussed in previous sections for the particular cases of droplet, bubbles, and particles, a classic sharp interface treatment supplemented with simplifying geometric assumptions led to the prediction of a unique (stable) thermodynamic equilibrium state corresponding to the global minimum of the system energy. The value of the contact angle at the energy minimum was given by Young's law for the cases of "ideal" interfaces that are perfectly spherical or flat. Underlying these classical results is the assumption that the studied interfaces are ideally smooth (i.e., zero r.m.s. roughness) and homogeneous (i.e., constant interfacial energies). A fundamental phenomenon observed for "real" surfaces with physico-chemical features of nano- and/or microscopic dimensions is the existence of a range of contact angle values for which different equilibrium conditions are observed [Figure 4a]. This phenomenon is known as contact angle hysteresis and has been extensively documented and studied in the literature.[14, 42, 43, 51-54] Moreover, certain surface features and chemical heterogeneities can lead to coexistent equilibrium states known as the Cassie-Baxter and Wenzel states, [55-58] where a liquid-fluid interface is either suspended over the surface features or collapsed onto the solid, as illustrated in Figure 4b. The described phenomena induced by interfacial heterogeneities and roughness indicate that colloidal and multiphase systems with "real" interfaces are metastable in nature and the observed multiple equilibrium conditions must correspond to local minima or saddle points in a topologically complex energy landscape of the system.[59-62]

Figure 4: Static contact angle hysteresis. (a) "Real" surfaces having physico-chemical heterogeneities exhibit a range of equilibrium contact angles. (b) "Rough" surfaces with macro- or microscale features exhibit coexisting metastable states known as Cassie-Baxter (suspended interface) and Wenzel (collapsed interface) wetting states.

## Thermodynamic Metastability

Modeling thermodynamic systems as composed by homogeneous phases separated by interfaces that are sharp and ideally smooth (e.g., perfectly flat or spherical) one arrives to thermodynamic potentials (e.g., eqn 8 and eqn 11) with a unique equilibrium state corresponding to the (global) energy minimum, for which $\theta = \theta_Y$. In order to consider the presence of heterogeneities and roughness in "real" surfaces it is customary to assume there is range of equilibrium contact angles $\theta_E^{min} \leq \theta \leq \theta_E^{max}$ and corresponding system variables (e.g., droplet radius, particle position) for which metastable equilibrium states are observed. Albeit effective to characterize equilibrium behavior, such approach is insufficient to describe non-equilibrium effects induced by the metastability of the system. To describe macroscopic non-equilibrium behaviors induced by very small heterogeneities one can include local minima in the energy potentials derived for ideally smooth interfaces. As elaborated in the next section, this approach can account for non-trivial non-equilibrium behaviors such as unexpectedly slow thermally activated relaxation to equilibrium and crossovers between dynamic and kinetic regimes.[63-69]

## Nanoscale Surface Heterogeneities

As illustrated in Figure 5a, let us analyze the case of an open system of fixed length $L_1$, width $L_2$, and height $L_3$ where a sharp interface between two fluids is allowed to move along the $x_1$-direction over a flat solid surface densely populated by small "defects" with an average projected area $A_d \ll L_i^2$ and height $h$. For the case of physical defects, the shape and characteristic dimensions can be obtained from topographic images with nanoscale resolution via Atomic Force Microscopy (AFM) [see Figure 5b] or other characterization techniques.

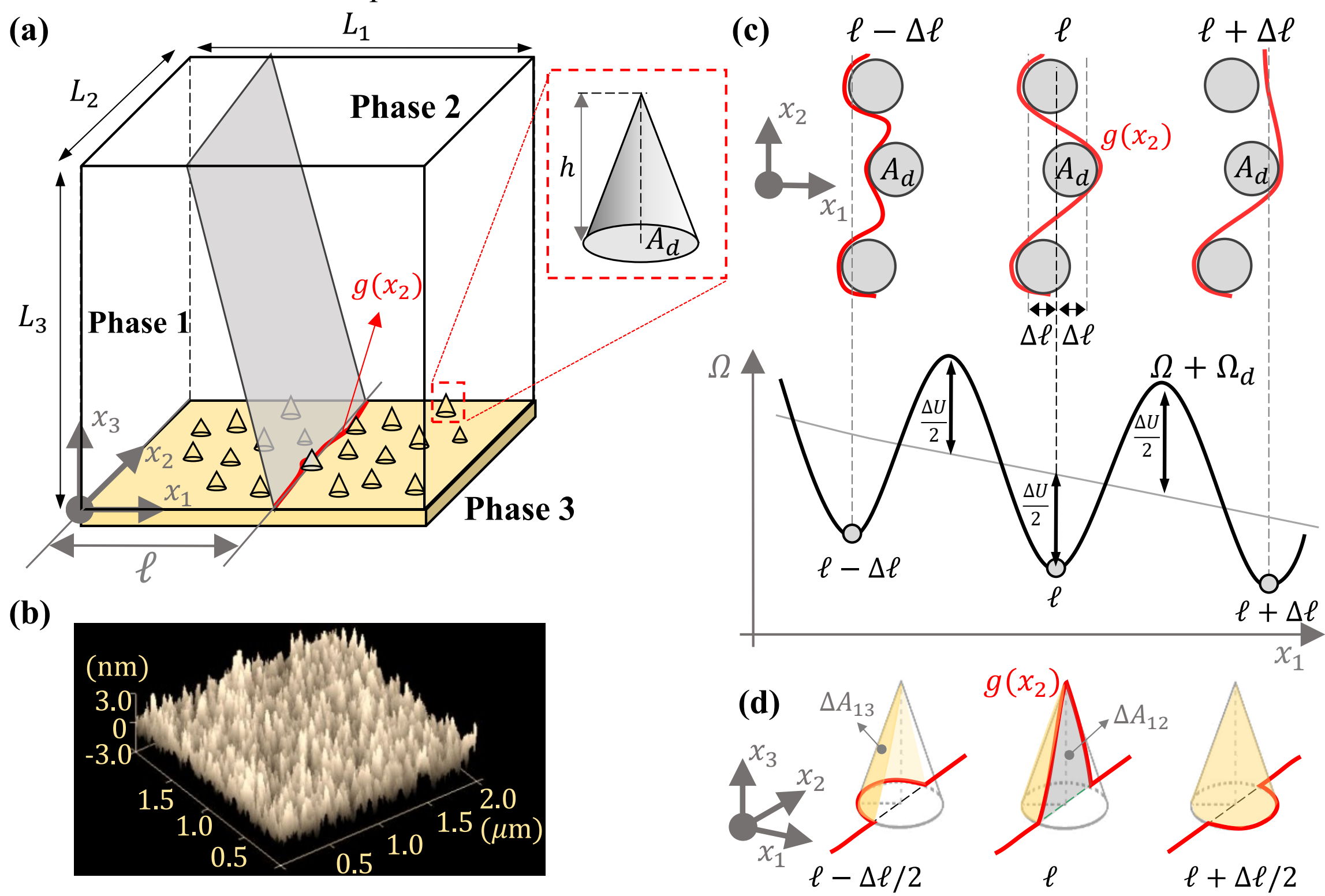

Figure 5: Nanoscale surface defects and metastability. (a) Wetting front at mean position $\ell$ along the $x_1$-direction on a surface populated with nanoscale defects. (b) AFM topographic image showing nanoscale surface features. (c) Mean contact line displacement and induced energy minima. (d) Modeled conical defect and variations in interfacial areas.

To construct simple expressions for thermodynamic potentials we will estimate the wetted surface areas as $A_{13} = \int_0^{L_2} g(x_2)\, dx_2$ and $A_{23} = L_1 L_2 - A_{13}$, where the function $g(x_2)$ gives the local position of the (three-dimensional) contact line perimeter along the direction of motion [see Figure 5a]. Following this approximation for the wetted surface areas it is convenient to introduce the mean contact line position $\ell = A_{13}/L_2$ to parametrize changes in the energy of the system. As illustrated in Figure 5b, when the contact line moves over a single nanoscale defect its average position $\ell$ increases by a small amount $\Delta\ell = A_d/L_2$. In addition, we must consider that the displacement of the contact line over a three dimensional defect [cf. Figure 5b-c] can induce an energy perturbation (increase or decrease) of magnitude $\Delta U \sim \gamma_{12} A_d$. Hence, for the case of a system with macro- or microscale dimensions ($L_i \simeq 1$ to 10 $\mu$m) and surface "defects" of nanoscale dimensions ($\sqrt{A_d} \simeq 1$ to 10 nm) we will have extremely small periods ($\Delta\ell \simeq 0.1$ to 1000 pm) for the energy perturbations. Given the extremely large number of metastable states that would be observed even for very small macroscopic displacements of mean contact line position $\ell$, spatial variations in the grand thermodynamic potential $\Omega$ that would correspond to an ideal surface without defects [cf. Figure 5b] can be effectively modeled by an average single-mode perturbation [23, 64-66, 70, 71]

$$\Omega_d = \frac{1}{2}\Delta U \, sin\left(\frac{2\pi A_{13}}{A_d} + \varphi\right) \quad (16)$$

where $\varphi$ is an arbitrary phase that varies the position of the global minimum by small amount $\sim\Delta\ell$; hereafter we will use $\varphi = 0$ for simplicity. Note that eqn (16) neglects small changes in the volume of the fluid phases due to the extremely small volume of the modeled nanoscale defects.

**Energy Barriers due to Nanoscale Surface Heterogeneities.** The energy barrier $\Delta U$ in eqn (16) must account for (1) nanoscale chemical heterogeneities that produce small changes in the local surface energies $\gamma_{13}$ and $\gamma_{23}$, and (2) nanoscale physical features that induce small changes in the interfacial areas $A_{ij}(\ell)$. For the case of a chemical heterogeneity on a flat surface that changes the local Young contact angle by a small amount $\Delta\theta_Y \lesssim 20°$ one can estimate $\Delta U = \beta\gamma_{12}A_d$ where $\beta \simeq \sin\theta_Y \, \Delta\theta_Y$. In the case of three-dimensional physical features such as hemispherical bumps, cylinders, or cones one can readily estimate the energy fluctuation via geometric arguments. For example, for the case of a cone with base area $A_d$ and height $h$ [cf. Figure 5a] we can obtain $\Delta U = \beta\gamma_{12}A_d$ where $\beta \simeq (h/\sqrt{\pi A_d})|1 - (\pi/2)\cos\theta_Y\,|$ by considering small changes $\Delta A_{12}$ and $\Delta A_{13}$ in interfacial areas as the contact line moves over the defect [cf. Figure 5c]. In general, given the complexity of different combined phenomena involved in the wetting of nanoscale surface features, the exact value of the energy fluctuation $\Delta U = \beta\gamma_{12}A_d$ where $0 \leq \beta \leq 1$ can be treated as a free model parameter that can be determined by fitting results from experimental observations.[65, 66]

Employing different models such as the so-called Molecular Kinetic Theory of contact lines[72] or Kramers theory[73] for the escape rate of a metastable state, energy barrier magnitudes $\Delta U$ ranging from 1 to 100 $k_BT$ have been reported to account for experimental observations in diverse multiphase systems, such as droplet spreading, microparticle adsorption at liquid interfaces, or drainage of microcapillaries.[63-66] It is worth noticing here that energy barriers larger than $\Delta U > 10k_BT$ can be produced by 1-nm defects ($A_d \sim 10^{-18}$ m$^2$) in water/air or water/oil systems for which $\gamma_{12} \sim 10^{-2}$ N/m.

# Kinetics and Dynamics at Interfaces

The kinetic and dynamic models discussed in this section can be employed to describe the time evolution of a finite set of $N_q$ macroscopic (observable) variables $\mathbf{q} = q_n(t)$ $(1 = 1, N_q)$ for a wide variety of systems involving liquid-fluid and fluid-solid interfaces. Despite the fact that the models in this section have demonstrated to be versatile and effective, they have significant limitations. Hence, when employing kinetic and dynamic models discussed in this section one must pay special attention to the physical conditions under which fundamental assumptions are valid, and model predictions can be accurate.

**The Markovian assumption.** It is worth remarking that a fundamental assumption underlying the models presented in this section is that future states of the system are predicted from a knowledge of the state variables at present time. This is equivalent to considering the system evolution is a Markovian process[74] where the "history" of previous dynamic events does not affect the future. In practice, this assumption allows us to model path-dependent dissipative work by using damping forces with the general form $f_n^d(t) = -\xi_n\dot{q}_n$, where $\xi_n$ is a configuration-dependent damping coefficient. Fundamentally, the Markovian assumption limits the applicability of the presented dynamic models to macroscopic processes much slower than microscopic relaxation processes bringing the system to local thermodynamic equilibrium, where physical parameters such as the interfacial tension $\gamma$ or fluid viscosity $\eta$ are well-characterized fluid properties. Fortunately, microscopic relaxation processes can be several orders of magnitude faster than macroscopic interfacial processes for numerous systems of technical interest. For example, simple molecular liquids at room temperature $T \simeq 300$ K have relaxation times $\tau_m = \sigma\sqrt{m/k_B T} = 1$ to 10 ps (here, $\sigma$ is the molecular diameter, $m$ the molecular mass, and $k_B T$ is the themal energy), while even for nanoscale systems (i.e., characteristic lengths $\ell =$ 10 to 100 nm) the characteristic time for dynamic wetting processes $\tau_M = \ell\eta/\gamma = 0.1$ to 1 ns can be one to three orders of magnitude larger.

**Dynamics vs. Kinetics.** State variables in a thermodynamic system with finite temperature $T$ experience fluctuations due to the "random" thermal motion of molecules composing the system. Typically, the energy of these fluctuations can be characterized by the thermal energy $k_B T$ (here, $k_B$ is the Boltzmann constant) and the intensity of equilibrium fluctuation of a variable $q_n$ is determined by the shape (e.g., well depth and curvature) of the system energy landscape. By assuming "ideal" interfaces the energy profiles determined in the previous sections (e.g., eqn 8 and eqn 11) do not present local minima. Systems with such "smooth" energy profiles are stable and exhibit a monotonic relaxation to thermodynamic equilibrium for fluctuations of arbitrary intensity. Moreover, the average evolution of state variables toward equilibrium can be effectively modeled by dynamic equations considering deterministic driving and damping forces (e.g., capillary forces, hydrodynamic drag). As elaborated in the previous section, the presence of nanoscale local heterogeneities and/or surface roughness in "real" surfaces, however, can produce energy profiles that are densely populated by local minima (Figure 5b) that correspond to metastable equilibrium states. For such metastable systems, the value of a state variable can "hop" between local minima in response to random thermal fluctuations. This thermally activated process can result in a nontrivial quasi-static evolution described by kinetic equations. The interplay between local minima induced by nanoscale defects and thermal motion can dominate the non-equilibrium dynamics and kinetics of relaxation of colloidal and multiphase systems as elaborated in the following sections.

### Lagrangian Mechanics and Deterministic Dynamics

In the framework of Lagrangian mechanics,[75] given a set of time-dependent state variables or "generalized coordinates" where $\mathbf{q} = q_n(t)$ $(n = 1, N_q)$ and $\dot{\mathbf{q}} = dq_n/dt$, one can define the system Lagrangian $\mathcal{L}(\mathbf{q}, \dot{\mathbf{q}}, t) = K - P$ where $K(\mathbf{q}, \dot{\mathbf{q}})$ and $P(\mathbf{q})$ and the total kinetic and potential energy, respectively. For conservative systems one can obtain equations for the evolution of each generalized coordinate from the Euler-Lagrange equations $\frac{\partial}{\partial t}(\partial\mathcal{L}/\partial\dot{q}_n) - \partial\mathcal{L}/\partial q_n = 0$. The Lagrangian mechanics approach can be extended to non-conservative system by means of the Rayleigh dissipation function $D(\mathbf{q}, \dot{\mathbf{q}}) = (1/2)c_{mn}\dot{q}_m\dot{q}_n$, where $c_{mn}(\mathbf{q})$ are coupling coefficients depending on the system configuration at a given time $t$. The Rayleigh dissipation function $D = \dot{E}/2$ is half the total energy dissipation rate $\dot{E}(t)$, which usually can be estimated via hydrodynamic equations, fluctuation-dissipation relations, or different physical arguments.[76] From the dissipation function $D$ one can obtain the (non-conservative) damping forces $f_n^d = -\partial D/\partial\dot{q}_n = -\xi_n\dot{q}_n$ where $\xi_n = c_{nm}\dot{q}_m$ is the effective damping coefficient determined by configuration-dependent coupling coefficients $c_{nm}(\mathbf{q})$. Hence, for a non-conservative system the evolution of state variable $q_n$ is formally described by the dynamic equation

$$\left(\frac{\partial}{\partial t}\frac{\partial}{\partial\dot{q}_n} - \frac{\partial}{\partial q_n}\right)K = f_n^c - \xi_n\dot{q}_n, \quad (17)$$

where the left hand side accounts for inertial effects, and the conservative forces $f_n^c = -\partial P/\partial q_n$ are given by the configuration dependent potential energy $P(\mathbf{q})$. The Markovian assumption discussed at the beginning of this section is invoked in the formulation of the energy dissipation $\dot{E}(t) = 2D$ in terms of instantaneous coupling coefficients $c_{nm}$.

**Langevin Dynamics and Thermal Fluctuations**

State variables in thermodynamic systems with finite temperature are expected to fluctuate as a result of thermal motion. This microscopic phenomenon gives rise to Brownian motion and mass diffusivity in the case of small particles immersed in a macroscopically quiescent fluid, as determined in the seminal works by A. Einstein[77] and M. Smoluchowski.[78] To model the dynamics of a Brownian particle, P. Langevin proposed in 1908 a stochastic ordinary differential equation that besides deterministic inertial and damping terms included and additional random force attributed to momentum and energy exchange with the surrounding fluid molecules.[79] Similar ideas have been extensively adopted to describe the evolution of collective variables in molecular systems, reaction coordinates in chemical kinetics, and order parameters in phase field models. For a rigorous derivation of generalized Langevin equations the reader is referred to the work by R. Zwanzig[79, 80] and H. Mori.[81] Here we adopt the essential ideas behind Langevin dynamics and include a stochastic term in eqn (17) to model the effect of thermal fluctuations, which leads to

$$\left(\frac{\partial}{\partial t}\frac{\partial}{\partial \dot{q}_n} - \frac{\partial}{\partial q_n}\right)K = f_n^c - \xi_n \dot{q}_n + \sqrt{2k_B T \xi_n}\, f(t), \quad (18)$$

where $f(t)$ is spatially uncorrelated Gaussian noise with $\langle f(t)\rangle = 0$ and $\langle f(t)f(t')\rangle = \delta(t-t')$ (hereafter the brackets $\langle\ \ \rangle$ indicate ensemble average). A crucial element in the construction of the Langevin equation for $q_n(t)$ (eqn 18) is that the magnitude of the stochastic term $\sqrt{2k_B T \xi_n}$ satisfies the so-called fluctuation-dissipation relation[79, 82] enforcing that, in the long time limit, the energy input from the modeled stochastic force is equal to the energy dissipated by damping forces.

For a thermodynamic system that evolves at constant volume $V$ and temperature $T$, changes in the Helmholtz free energy $F$ (eqn 2) and grand potential $\Omega$ (eqn 3) give the reversible work performed by a closed or open system, respectively. We can thus specialize eqn (17) and eqn (18) for the particular thermodynamic systems studied in previous sections. In addition, we will consider the effect of nanoscale heterogeneities in the system energy as modeled by eqn (16).

**Hemispherical Droplet on a Solid Surface**

For the case of small hemispherical droplets spreading on a flat surface [See Figure 2b], changes in the grand potential are given by eqn (9) and we can adopt the contact radius $R$ as a generalized coordinate. Conservative forces in eqn (18) are thus given by $f^c = -d(\Omega + \Omega_d)/dR$ after incorporating eqn (16) in order to consider nanoscale defects of area $A_d$. Although fluid flow produces a finite kinetic energy $K(R,\dot{R})$, for small Reynolds numbers $Re = \rho_1 \dot{R} R/\eta_1 \ll 1$ (here, $\rho_1$ and $\eta_1$ are the mass density and viscosity of the liquid phase) dissipative forces will dominate and thus $K/R \ll \xi_R \dot{R}$, where $\xi_R$ is the effective damping coefficient. Under the described assumptions, eqn (18) gives

$$\xi_R \dot{R} = 2\pi R \gamma_{12}(\cos\theta(R) - \cos\theta_Y) + f_R + \sqrt{2k_B T \xi_R}\, f(t), \quad (19)$$

where

$$f_R = -\frac{d\Omega_d}{dR} = -\Delta U \frac{2\pi^2 R}{A_d} \cos\left(\frac{2\pi^2 R^2}{A_d}\right) \quad (20)$$

is a conservative and deterministic force modeling the average effect of nanoscale surface defects on the spreading dynamics. The damping coefficient $\xi_R(R)$ in eqn (19) can be estimated by considering various mechanisms that contribute to the energy dissipation rate during the dynamic spreading process. Considering solely hydrodynamic effects, the damping coefficient can be determined by the Voinov-Cox model [83, 84] or lubrication theory for the case of thin droplets with low contact angles [85]. Additional physical processes such as irreversible adsorption-desorption of fluid molecules at the solid surface have been modeled via the MKT models [72] that will be described in a later section.

### Spherical Particle at a Flat Interface

For the case of a spherical particle or radius R with nanoscale surface defects of characteristic area $A_d$ that straddles a flat liquid-fluid interface at position z = 0 [See Figure 2c]. Combining eqn (18) with eqn (11) and eqn (16) for the grand potential leads to

$$(m_p + m_f)\ddot{z} = -\xi_z\dot{z} - 2\pi\gamma_{12}(z - z_E) + f_z + \sqrt{2k_BT\xi_z}\, f(t) \quad (21)$$

where $m_p$ is the particle mass, $m_f$ is the added mass due to fluid moving with the particle, $\xi_z$ is the effective damping coefficient, and

$$f_z = -\frac{d\Omega_d}{dz} = -\Delta U\frac{2\pi R}{A_d}cos\left(\frac{2\pi Rz}{A_d}\right) \quad (22)$$

is the force induced by nanoscale surface defects when the particle moves normal to the interface. As for the case of droplet spreading, the damping coefficient $\xi_z(z)$ can be obtained by considering dissipation due to hydrodynamic effects and/or adsorption-desorption of molecules at the contact line. In addition, recent works have proposed that the damping coefficient $\xi_z$ must account for random thermal fluctuations of the contact line, which can be accomplished via Green-Kubo relations involving the time autocorrelation of fluctuating surface forces.[66, 86] The Langevin model in eqns (21)-(22) has been recently employed to account for experimental observations for microparticles with different surface functionalization at a water-oil interface by employing defects areas $A_d \simeq 10$ to 90 $nm^2$ and energy barriers $\Delta U = 20$ to 360 $k_BT$.[66]

## Regime Crossovers

It is worth remarking that evolution equations derived from eqn (19) can equally model: (1) dynamic regimes dominated by deterministic forces when the system is far from equilibrium and $|\partial P/\partial q_n| \gg |\sqrt{2k_BT\xi_n}|$, and (2) kinetic regimes dominated by random thermally activated processes when the system is sufficiently close to equilibrium and $\partial P/\partial q_n \to 0$. Near thermodynamic equilibrium conditions, the thermodynamic potentials in eqns (1)-(3) can be approximated by second-order Taylor expansion and conservative are approximately linear.

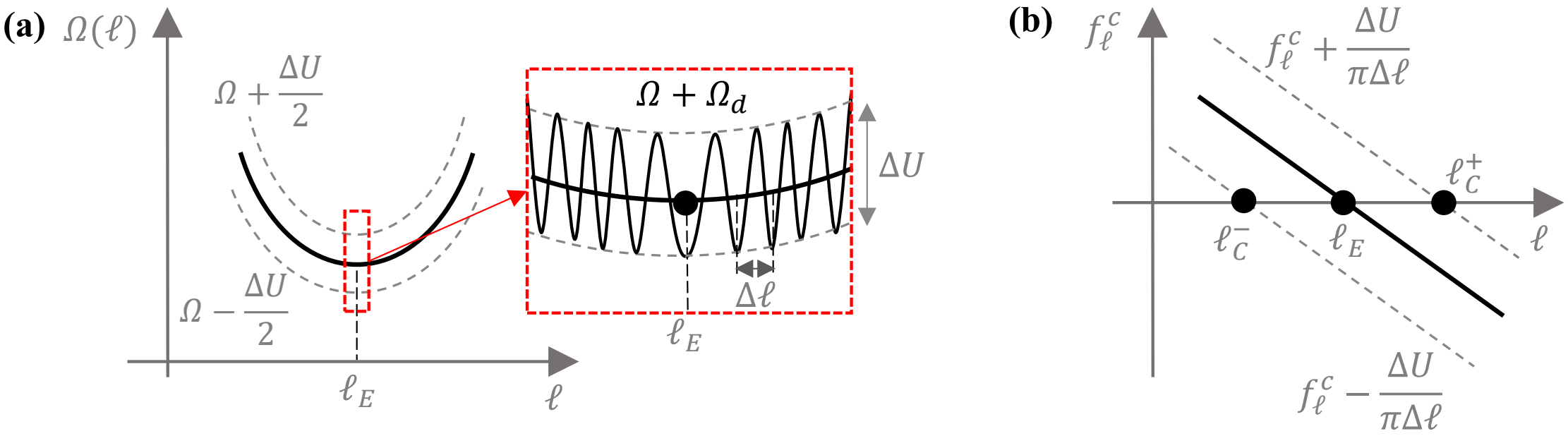

Figure 6: Crossover to kinetic regimes dominated by thermally activated transitions between metastable states. (a) Grand potential profile $\Omega(\ell)$ considering energy barriers $\Delta U$ induced by nanoscale surface defects. (b) Conservative driving forces $f_\ell^c$ and crossover points $\ell_C^+$ and $\ell_C^+$ from "above" or "below" (+/-) toward the expected equilibrium at $\ell_E$.

The described crossover between dynamic and kinetic regimes takes place around a critical value $q_n = q_C$ for which local minima at which $\partial(\Omega + \Omega_d)/\partial q_n = 0$ begin to appear. Considering once again an open system with constant volume and temperature for which reversible work is given by changes in the grand potential $\Omega$, we find that conservative forces are $f_n^c = -\partial P/\partial q_n = -\partial\Omega/\partial q_n$. As illustrated in Figure 6a-b, for the case that $\Omega = \Omega(\ell)$ can be parametrized by the average position of the contact line $\ell$, we find that conservative forces near equilibrium are $f_\ell^c \simeq k_E(\ell - \ell_E)^2$ where $\ell_E$ is the equilibrium value of the state variable $\ell$, and $k_E = d^2\Omega(\ell_E)/d\ell^2$. Adopting the single-mode perturbation in eqn (16) with magnitude $\Delta U$ and period $\Delta\ell = A_d/L_2$ we can estimate the effective force $-\partial\Omega_d/\partial\ell$ due to nanoscale defects. We can thus find that the crossover point $\ell_C$ is given by

$$|\ell_C - \ell_E| \simeq \frac{1}{k_E}\frac{\pi\Delta U}{\Delta\ell}. \quad (23)$$

When a regime crossover is experimentally observed at $\ell_C$, eqn (23) can be employed to estimate the period $\Delta\ell$ and infer the characteristic area of surface defects $A_d = L_2\Delta\ell$, where $L_2$ is the average contact line perimeter.[66, 87]

## Kramers Theory for Wetting: Thermally Activated Escape from Metastable States

When a thermodynamic system is sufficiently close to equilibrium, nanoscale heterogeneities are likely to induce numerous local minima and the time evolution of the system becomes a series of thermally activated transitions or "hops" between metastable states. Under such conditions the evolution of a state variable $q_n$ governed by eqn (18) can be effectively described by rate equations according to the celebrated Kramers theory of thermally activated escape rates.[73, 88] As before, let us analyze the prototypical case of an open system for which the thermodynamic potentials including nanoscale surface defects $\Omega'(\ell) = \Omega + \Omega_d$ can be parametrized by the average contact line position $\ell$. Let us also assume that according to eqn (16) nanoscale surface defects induce metastable states separated by a characteristic energy barrier $\Delta U$ and occur with a period $\Delta\ell \ll \ell$. Given a series of local minima at position $\ell_o$ and neighboring maxima at $\ell_\pm = \ell_o \pm \Delta\ell/2$, the forward/backward $(+/-)$ escape rates for $|\ell - \ell_o| < \Delta\ell/2$ given by Kramers theory are

$$\Gamma_\pm(\ell) = \frac{1}{2\pi\xi}\sqrt{\frac{\partial^2\Omega'_o}{\partial\ell^2}\left|\frac{\partial^2\Omega'_\pm}{\partial\ell^2}\right|}\, exp\left(-\frac{\Delta U}{k_B T}\right) exp\left(-\frac{\Omega_\pm - \Omega_o}{k_B T}\right), \quad (24)$$

where $\xi(\ell)$ is the effective damping coefficient, $\Omega'_o = \Omega'(\ell_o)$, and $\Omega'_\pm = \Omega'(\ell_\pm)$. The rate equation for the average evolution of $\ell$ is thus given by

$$\frac{d\ell}{dt} = \Delta\ell(\Gamma_+ - \Gamma_-) \quad (25)$$

where the period or "hop" length $\Delta\ell = A_d/L_2$ is given by the characteristic defect area $A_d$ and mean contact line perimeter $L_2$. For the particular case of near equilibrium conditions at $\ell = \ell_E$ and assuming a constant damping coefficient $\xi \simeq \xi(\ell_E)$ eqn (25) can be integrated analytically, which gives a nearly logarithmic in time evolution.[64]

Kramers theory for wetting (eqns 24-25) gives accurate results for nano- to mesoscale surface defects with projected areas $A_d = 1$ to 100 nm$^2$ when estimating the proper magnitude of energy barriers $\Delta U$ induced by the three-dimensional defect geometry. In particular, eqns 24-25 have successfully accounted for the experimentally observed near-equilibrium behavior of single microparticles adsorbed at a liquid interface[66] and shear-driven drainage of microcapillaries.[65]

### Molecular Kinetic Theory for Dynamic Contact Lines

This section describes the so-called Molecular Kinetic Theory (MKT) proposed by T.D. Blake and co-workers[72] to describe the non-equilibrium dynamics of contact lines by considering the kinetics of molecular adsorption and desorption. As the three-phase contact line illustrated in Figure 5a moves along the $x_1$-direction and the interfacial area $A_{13}$ increases, molecules composing the fluid in phase 1 must become in contact with the solid surface at "adsorption sites" of characteristic size $\lambda_M$ that were previously occupied by molecules composing fluid phase 2. Building on Eyring's theory of chemical kinetics, MKT estimates the equilibrium rate for the adsorption-desorption process as[72, 89]

$$\Gamma_0 = \left(\frac{k_B T}{\eta \nu}\right) exp\left(\frac{W_a}{k_B T}\right) \quad (26)$$

where $\eta$ is the adopted characteristic viscosity (e.g., the viscosity of the most viscous fluid), $\nu$ is the characteristic molecular volume, and $W_a = \lambda_M^2 \, \gamma_{12}(1 + \cos\theta_Y)$ is the so-called work of adhesion. When the system is out equilibrium and the observed contact angle is $\theta \neq \theta_Y$, the contact line will experience a net driving force per unit-length $f_b \simeq \gamma_{12}(\cos\theta - \cos\theta_Y)$ that produces an irreversible work $W_b \simeq \lambda_M^2 \gamma_{12}(\cos\theta - \cos\theta_Y)$ when moving over an adsorption site. Following MKT, the average displacement rate can be described by a rate equation $d\ell/dt = \lambda_M(\Gamma_+ - \Gamma_-)$ where the "forward/backward" $(+/-)$ adsorption rates are

$$\Gamma_{\pm} = \Gamma_0 \, exp(\mp W_b) = \left(\frac{k_B T}{\eta \nu}\right) exp\left(\frac{W_a}{k_B T}\right) exp\left(\mp \frac{\lambda_M^2 \gamma_{12}(cos\,\theta - cos\,\theta_Y)}{2 k_B T}\right). \quad (27)$$

For the particular case that $W_b \ll 2k_B T$ and an overdamped system for which $-\xi d\ell/dt = f_b$, the kinetic equation proposed by MKT can be cast as a dynamic equation

$$\frac{d\ell}{dt} = \lambda_M(\Gamma_+ + \Gamma_-) \simeq \frac{1}{\xi} L_2 \gamma_{12}(cos\,\theta - cos\,\theta_Y) \quad (28)$$

for a contact line of perimeter $L_2$ where the effective damping coefficient is

$$\xi = L_2 \left(\frac{\eta \nu}{\lambda_M^3}\right) exp\left(-\frac{\lambda_M^2 \, \gamma_{12}(1 + cos\,\theta_Y)}{k_B T}\right). \quad (29)$$

In the MKT approach, the energy "consumed" in the adsorption-desorption of liquid molecules when the contact line moves at speed $d\ell/dt$ is modeled as a linear dissipative force $f_d = -\xi d\ell/dt$ determined by the damping coefficient in eqn (29). Analytical fits employing the MKT model have been reported to account for experimental observations in different systems for adsorption site sizes $\lambda_M =$ 0.3 to 5 nm.[90] To be consistent with the MKT model assumptions, the adsorption site must be comparable to the size of fluid molecules ($\lambda_M \simeq \sqrt[3]{\nu}$) and the system must be sufficiently close to thermodynamic equilibrium so that $\cos\theta - \cos\theta_Y \ll 2k_B T/\lambda_M^2 \gamma_{12}$ and the linearization of eqn (28) that leads to the damping coefficient in eqn (29) is valid.

## Future Directions

Thermodynamic models currently employed to study colloidal and multiphase systems have been developed over two centuries ago to tackle the challenge of describing macroscopic systems essential to modern technologies developed in the industrial revolution of the 1800s. With the advent of nanotechnology and advanced characterization techniques in the 21st century, we now have the ability to synthesize and control truly nanoscopic systems. A current theoretical challenge is to extend classical continuum-based descriptions for the accurate analysis of system with dimensions approaching the molecular scale. Work in this direction must consider the complex interplay between thermal motion, finite range molecular interactions, and nanoscale surface heterogeneities of natural (random) or synthetic (ordered) nature. Moreover, physical systems such as complex fluids or colloids exhibit a finite relaxation time in response to a perturbation or external actuation, and it is not uncommon to observe strong non-Markovian effects. While all these effects are conceptually and mathematically difficult to model, they open the opportunity to exciting technological developments. For example, nanostructured surfaces can be synthesized with ordered and precise geometric features of the order of 10 nm via self-assembly techniques.[91-94] These structures could be designed to control the energy barriers and periods of the induced metastable states and thus control the kinetics of wetting by different fluids and droplets of different sizes. Moreover, asymmetric nanostructures could be designed to produce asymmetric energy barriers and rectify random thermal fluctuations into directional interfacial motion,[95-98] in similar fashion to a Brownian ratchet.[99] Non-Markovian effects and the associated correlated thermal motion[100, 101] could be exploited to amplify the response to external actuation with a given frequency, which resembles the phenomenon of stochastic resonance,[102, 103] and effectively drive colloidal systems away from undesired metastable configurations. Further development and application of thermodynamic, dynamics, and kinetic models described in this article can guide the study of the phenomena discussed in this section, among many other nontrivial phenomena, and open concrete opportunities for new technological developments involving colloidal and multiphase systems.